\pdfoutput=1

\PassOptionsToPackage{table,dvipsnames}{xcolor} 

\documentclass[sigconf]{acmart}

\setcopyright{none}
\renewcommand\footnotetextcopyrightpermission[1]{}

\usepackage{enumitem}
\usepackage[table]{xcolor}
\usepackage{tikz}
\usetikzlibrary{shapes.geometric} 
\usepackage{booktabs}
\usepackage{amsmath}
\usepackage[most]{tcolorbox}
\usepackage{tabularx}
\usepackage{todonotes}
\presetkeys{todonotes}{inline}{}

\newcommand{\github}{\url{https://anonymous.4open.science/r/rag-redunancy-C8BC}}

\AtBeginDocument{%
  }

\begin{document}

\title[How retriever redundancy and diversity impact RAG effectiveness]{How retriever redundancy and diversity \\ impact RAG effectiveness}

\author{Jonathan J Ross}
\correspondingauthor
\email{jonathan.ross@uq.edu.au}
\affiliation{%
	\institution{The University of Queensland }
	\city{Brisbane}
	\state{Queensland}
	\country{Australia}
}
\author{Bevan Koopman}
\email{bevan.koopman@csiro.au}
\affiliation{%
	\institution{CSIRO / The University of Queensland }
	\city{Brisbane}
	\state{Queensland}
	\country{Australia}
}
\author{Anton van der Vegt}
\email{a.vandervegt@uq.edu.au}
\affiliation{%
	\institution{The University of Queensland }
	\city{Brisbane}
	\state{Queensland}
	\country{Australia}
}
\author{Guido Zuccon}
\email{g.zuccon@uq.edu.au}
\affiliation{%
	\institution{The University of Queensland }
	\city{Brisbane}
	\state{Queensland}
	\country{Australia}
}

\renewcommand{\shortauthors}{Ross et al.}

\begin{abstract}

In RAG, while the retriever typically ranks documents by their individual relevance to the query, the generator instead produces an answer based on the retrieved documents as a whole.  This paper investigates how redundancy and diversity from the retrieved document set impact the generator in terms of answer correctness.

Previous work has provided a mix of findings: some showing that redundancy improves generation by reinforcing relevant information, others that LLM-based paraphrasing of the same content may be beneficial. Many of these studies did not control for confounding factors like whether the documents contained the exact answer or not, and if parametric knowledge plays a role. We conduct a carefully controlled experiment investigating three key scenarios of retrieved document sets: 1) Duplicate (exact copies of the same document), 2) Paraphrased (LLM rephrased versions of one document) and 3) Diverse (documents from different genres each containing relevant information in different forms). We control for which documents contain the answer in exact match or rephrased form.

Evaluation is done with FictionalQA, a synthetic, fictional question-answer dataset that ensures the LLM generator prior knowledge cannot answer the question; the answer must come from retrieved documents. We show that duplicate redundancy and LLM paraphrasing does not significantly improve answer correctness. However, providing diverse documents is highly beneficial, improving answer correctness by 17\%--47\%. We further show this improvement is driven by diverse forms of document genre (news, blogs, etc.) alone and not a consequence of more relevant answer being available to generator. Our findings help to direct more attention to how new retrieval methods might improve RAG by catering to the generator preference for diversity in retrieval results.
Code and results are provided in \github.
\end{abstract}


\keywords{retrieval-augmented generation, retrieval redundancy, result
  diversity, question answering, large language models}

\maketitle

\pagestyle{plain}


%

\section{Introduction}

Retrieval augmented generation (RAG) improves a generator's output by drawing on external documents rather than relying only on what the large language model (LLM) learned in training \citep{lewis_retrieval-augmented_2020}. The retriever typically scores each document individually for its relevance to the query \citep{karpukhin_dpr_2020}. However, the generator receives these documents together, and collectively they have properties that affect generation \citep{liu_lost_2024, cuconasu_power_2024}

One such collective property is redundancy. Among retrieved documents, content to help the generator correctly answer the question may turn up in multiple documents, expressed in the same or different words.  Because retrieval cannot see which documents actually lead to a correct answer, it falls back on similarity in wording or meaning as a proxy. A family of methods -- diversity-aware reranking, coverage-oriented selection, subset selection \citep{carbonell_use_1998,santos2012role,zuccon2012top,santos2015search,ju_controlled_2025,lee_shifting_2025,ju2026search} -- uses this proxy to steer the retrieved set toward variety, scoring a document lower when it overlaps semantically with what has already been selected. The premise is that redundancy is wasteful under a fixed retrieval budget, and that similarity is a reasonable stand-in for redundancy when true redundancy itself cannot be observed. Avoiding redundancy is driven by the requirement that human users don't want to see the same information presented to them again.

In RAG, the output of the retriever is not for a human but a generator; here redundancy may not be wasteful. LLMs do not simply check whether a fact is present; they condition on the full distribution of input text from the retriever. Repeated information may, therefore, change the perceived frequency, salience, or reliability of a claim, even when it adds no new factual content. It may reinforce the correct answer or clarify a weakly phrased passage; it may also crowd out complementary evidence, inflate confidence, or encourage the model to treat repetition as evidential strength. This issue is particularly relevant for short-form question answering, where the answer is a single fact and extra copies -- identical or paraphrased -- add no obviously new information. Whether they nonetheless change what the model does is an open question, and motivates this study.

Previous work has not fully reconciled the impact of redundancy and diversity in retrieval results across different generation settings. In macro-level report generation, maximising information coverage while penalising redundant documents strongly correlates with higher downstream generation quality~\cite{samuel2026beyond}. However, in short-form QA, results are more varied. Some studies conclude redundancy should be minimised \citep{ju_controlled_2025}. 
Other studies show that duplicating a retrieved passage verbatim does not significantly change answer correctness but does change the model's attribution of its answer~\citep{jain_cue-r_2026}.
Regenerating multiple, paraphrased versions of a single document and presenting these to a generator has also shown to influence generator correctness more than pure duplicated redundancy~\citep{naphade_rational_2026}. Finally, repetition from a selected document source (e.g., government or newspaper sources) seems to drive answer correctness more than a diverse set of sources.

This study looks deeper than the basic question of "does redundant/diverse retrieval results affect generator correctness"; it teases out which aspect of redundancy the generator responds to: the recurrence of the information itself, the repetition of its structure and wording, or the number of diverse sources attesting it. It is a controlled mechanistic study, isolating what redundancy alone does to the generator, motivated by, but not resolving, whether retrieval should minimise redundancy. We use FictionalQA \citep{kirchenbauer_fictionalqa_2026}, a synthetic dataset whose target facts concern fictional events, meaning the answer to questions cannot come from model training (parametric knowledge) and must come from retrieved passages. This removes the parametric prior as an alternative source -- the confound most work leaves uncontrolled -- and it matters most for confidence: an output probability over a real fact blends parametric certainty with whatever the retrieved set contributes, and the two cannot be separated; over a fictional fact, the probability is attributable to the retrieved evidence alone.  Holding relevant retriever content fixed, we vary how many times it recurs and the form of each recurrence (duplicate copies, paraphrased copies, or documents from diverse sources) -- separating repetition from the form it takes, not necessarily mimicking what real retrievers produce. Because the answer-bearing information stays constant, any change in output traces to redundancy rather than new content. We measure answer correctness as the final outcomes. 
Holding the relevant information provided to the generator as fixed and varying only its recurrence isolates redundancy from content, letting us measure what redundancy alone does to a generator's output.

\section{Related Work}

Two primary lines of work bear on this question without resolving it: retrieval-coverage evaluation and behavioral perturbation studies under evidence variation.

\paragraph{Redundancy in retrieval and coverage.}
The assumption that humans do not want to see redundant information and therefore that retrievers should minimise redundancy has carried into RAG through coverage-based evaluation~\cite{ju_controlled_2025,samuel2026coveragebench}. For example, CRUX \citep{ju_controlled_2025} measures retrieval quality by how many diverse sub-questions a retrieved set can answer, penalizing redundant passages for ``crowding out'' other relevant material. However, this logic implicitly assumes there is other material to crowd out—it applies neatly to long-form responses spanning multiple query aspects, but breaks down in short-form QA where a single document contains the complete fact. It also ignores the possibility of redundant information further reinforcing a correct answer by increasing generate token probabilities.

\paragraph{LLM sensitivity to evidence and credibility cues.}
Research on how models weigh evidence shows that generator behavior depends heavily on presentation and query relevance rather than source authority. \citet{wan_evidence_2024} demonstrate that under conflicting retrieved evidence, LLM answers are driven primarily by query relevance while largely ignoring human-oriented credibility cues such as the source of the document. However, their focus is on resolving active conflict among real-world documents for controversial or binary questions. In contrast, our setting holds the factual answer constant and varies only the source (in our case document genre), isolating whether repeated evidence changes correctness when no new factual content is added.

\paragraph{Perturbation studies and repetition.}
Perturbation studies further show that redundancy is not behaviorally inert, though existing work leaves critical gaps. CUE-R \citep{jain_cue-r_2026} duplicates a retrieved item verbatim and finds overall answer correctness unchanged, though the model's internal attribution shifts. GroupQA \citep{naphade_rational_2026} finds that paraphrased repetition can flip a model's belief more readily than the same number of diverse documents, treating repetition as consensus. Similarly, \citet{schuster_whose_2026} show that verbatim repetition from a single source sways models as strongly as a genuine multi-source preference, suggesting LLMs count token frequency rather than independent corroboration. 

\paragraph{The gap.}
None of these studies isolate the effect of pure structural redundancy on non-parametric fact extraction. CUE-R and GroupQA do not isolate redundancy from the model's parametric prior: CUE-R tests real-world facts where its closed-book setting answers roughly 20\% of the time without evidence, while GroupQA focuses on controversial claims with no ground truth. \citet{schuster_whose_2026} control for parametric priors using synthetic entities, but evaluate redundancy strictly under knowledge conflict (where repetition serves as a tiebreaker between contradictory claims) and restrict their repetition to duplicate copies. Short-form factual QA free of parametric priors and factual conflict thus remains untested: evaluating a single, uncontested fact varied systematically across frequency, form (duplication vs. paraphrased), and source provenance.
\section{Experimental Setup}

\begin{figure*}[t]
	\centering
	\includegraphics[width=\textwidth]{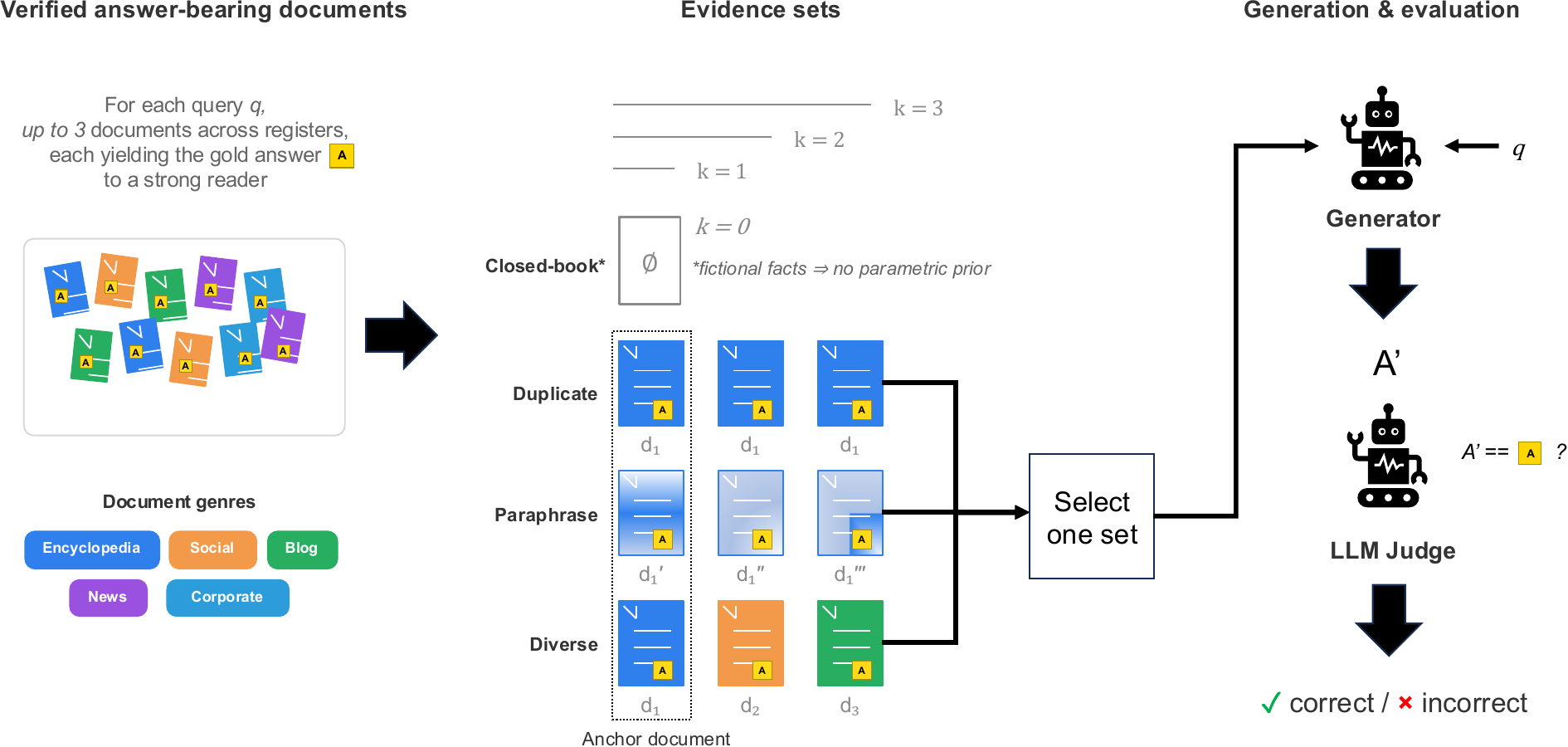}
	\caption{Overview of our experimental pipeline. \textbf{Left:} for each query $q$, we pool fictional documents verified to yield the gold answer $A$ Facts are fictional so generators have no parametric prior to draw on. Documents span five genres. \textbf{Middle:} evidence sets hold the answer fixed and vary only the number ($k=0$--$3$) and form of its recurrence: \emph{duplicate} ($k$ copies of the anchor document $d_1$), \emph{paraphrase} ($k$ rewordings of $d_1$), and \emph{diverse} (the first $k$ independent documents); a closed-book arm ($k=0$) provides the no-evidence floor. \emph{Duplicate} and \emph{diverse} are equivalent when $k=1$ (the anchor $d_1$ alone); the \emph{paraphrase} set contains only rewordings of $d_1$, so its $k=1$ set is a single reworded copy rather than the anchor verbatim. \textbf{Right:} each generator answers $q$ from a single evidence set, and an LLM judge scores its response $A'$ against the gold answer.} \label{fig:method_overview}
\end{figure*}

To test how redundancy and diversity affects generation, we use a short-answer QA task: a generator receives a query and $k = 0$--$5$ in context documents and produces a single answer. The answer is scored against a ``gold'' answer by an LLM judge (Figure~\ref{fig:method_overview}). 

Holding the query and gold answer fixed, every condition at context size $k$ supplies $k$ answer-supporting documents. The conditions vary three aspects of redundancy/diversity independently: 
\begin{enumerate}
	\item how often the answer recurs (the sweep over $k$);
	\item how the document is worded (duplicate copies of an anchor document versus paraphrases of it);
	\item where it comes from (paraphrases versus documents drawn from different genres).  
\end{enumerate}

A document counts as ``answer-supporting'' if a strong reader (GPT-4o) produces the gold answer from it alone.  

The duplicate and diverse conditions are the same when $k = 1$, where both present the anchor document alone; the paraphrase condition is anchored on the same document but presents a LLM-based rewording of it.  

Documents, queries, and answers come from FictionalQA \citep{kirchenbauer_fictionalqa_2026}, a synthetic dataset of fictional entities and events.  Since FictionalQA facts are made up, they do not appear in any pre-training corpus; thus, the generator cannot answer from parametric knowledge alone; this removes memorisation as an explanation for correct answers, so that performance differences across conditions are attributable purely to the documents passed as input.

\subsection{Dataset}


Our design places three requirements on the corpus. First, it must support short-answer QA: each query asks about a single fact and is scored against a fixed short answer, so that recurrence of the answer across documents is well-defined. Second, the facts must appear in no pre-training data, so that generators cannot answer from parametric knowledge and correctness is attributable entirely to the evidence in context. Third, each fact must be stated in several independently written documents, so that the diverse condition adds different sources rather than copies. The FictionalQA~\citep{kirchenbauer_fictionalqa_2026} dataset meets these requirements. It comprises: 1{,}500 documents covering 100 fictional events (15 documents per event), spanning five genres (news, social media, corporate, encyclopedia, blog); and 7{,}500 questions, each generated from a single source document together with a short gold answer and a declarative statement of the underlying fact. Because the events are fictional, the facts appear in no pre-training corpus; closed-book accuracy is 0.01--0.03 across all five generators, confirming the absence of a usable prior (Table~\ref{tab:main_correctness}).   From this dataset we take the documents, questions, answers, and a procedure for verifying document support for an answer; the next two subsections describe how we extract fact-level queries from the provided questions, and assemble sets of verified evidence for each query.

 \paragraph{Query construction} Of the 7,500 FictionalQA questions, many relating to the same corpus-based fact are similarly worded.  Consequently, we perform question de-duplication, which serves two purposes.  First, it prevents correlation between similarly worded questions related to the same fact. Second, it identifies documents that support the same de-duplicated question.  De-duplication itself was performed in three steps.  First, we remove the 959 questions whose source documents fail answerability verification: a strong reader given only the document cannot generate the gold answer (details below).  Second, questions were clustered by event, normalised fictional fact, and normalised answer.   These questions all target the same fact from the corpus and share the same answer string.  The resulting cluster identifies all the documents from which the questions were generated.  From the 6,541 remaining questions, 2,756 de-duplication clusters were created.  Each cluster---one fact, one gold answer, and the documents that support them---constitutes a \emph{query}.  Importantly, no cluster contains a document more than once.    Third, because the cluster contains question variants, one is selected as the query's \emph{canonical question}---the question posed in all conditions. For its selection, the cluster's members are randomly ordered and the first member's question is taken.  The selected member contributes the canonical question; the documents shown to generators are chosen separately (\S\ref{sec:evidence}).

The pool of 2,756 queries is filtered further based on the availability of evidence. Details on query filtering and evidence selection are provided in \S\ref{sec:evidence}.


\paragraph{Document verification.} \label{par:doc-verification} A document is verified as \emph{answer-supporting} for a query if a strong reader (GPT-4o), given the document alone, generates an answer graded as equivalent to the query's gold answer. Verification does not require the answer string to appear in the document---only that a reader can generate the answer from that document. In \S\ref{sec:answer-screened} we test whether our results depend on the answer string being present.  This replicates FictionalQA's informed-answerability grading with the same model (GPT-4o-2024-08-06), prompts, and two-step procedure; our regrade matches the released grades on 98.6\% of (query, document) pairs. Besides validating the released grades, regrading serves a second purpose: FictionalQA graded each document against the question it generated, whereas we pose the query's canonical question ---the question our generators see. Agreement is 98.7\% where the two questions coincide and 97.9\% where they differ, so the rewording changes almost no grades. The 959 questions discarded during query construction are those whose source document fails this check.

\subsection{Evidence sets}
\label{sec:evidence}

 For each query we assemble evidence sets under four conditions at context sizes $k \in \{0,1,2,3,4,5\}$, closed-book ($k=0$, no documents), duplicate ($k$ verbatim copies of the query's \emph{anchor document}---one designated verified document per query, selected as described below), paraphrase ($k$ independently LLM reworded copies of the anchor document), and diverse (the anchor document plus $k-1$ further verified documents). The duplicate and diverse conditions coincide at $k=1$, where both present the anchor document alone; the paraphrase condition presents a reworded copy of the anchor rather than the anchor itself, even at $k=1$.

 \paragraph{Anchor Document Selection} Documents fit within 5 genres: news, social media, corporate, encyclopedia and blogs. Each query's anchor document is chosen from its verified documents under a genre balance constraint: queries are assigned anchor genres so that each genre anchors roughly one fifth of all queries, and the anchor is then drawn at random from the query's verified documents of the assigned genre.
 
\paragraph{Duplicate Evidence Set} The duplicate setting contains $k$ duplicate copies of the same anchor document.

\paragraph{Paraphrase construction.}
Every document in a paraphrase evidence set is an independent paraphrase of the anchor document. Paraphrases are generated by Qwen2.5-32B-Instruct, using a temperature of 0.8 and an instruction to rephrase the document in its own words while preserving all information and adding none (Figure~\ref{fig:prompt-paraphrase}). Answer related content survives the paraphrase only if the model follows the instruction. Paraphrases are checked for diverseness: one that duplicates an existing paraphrase is re-generated so the $k$ paraphrases in an evidence set are pairwise diverse. The paraphrase evidence set at size $k$ presents the first $k$ of an anchor's five paraphrases, so raising $k$ adds a new paraphrase while holding the earlier ones fixed.

\paragraph{Diverse evidence.}  The diverse setting contains documents from different genres. Each query's verified documents are arranged once: the anchor document first, then one verified document from each remaining genre, in random genre order. The diverse evidence set at size $k$ presents the first $k$ documents of this order, so it always contains $k$ documents from $k$ different genres. A query therefore supports the diverse condition at size $k$ only if its verified documents span at least $k$ genres. 

\paragraph{Experimental sample}  The diverse condition limits which queries can be used; its evidence set at size $k$ needs verified documents from $k$ genres (recall that verified means that the document produces a correct answer when given to a generator).  Duplicate and paraphrase require only the anchor document, so every query supports them at every $k$. Verified documents span all five genres for 1,900 of the 2,756 queries (Table~\ref{tab:dataset-properties}). Our main analyses use these 1,900: every one of them supports every condition at every $k$, so differences between conditions cannot come from comparing different queries. The canonical question's source document may or may not appear in a given evidence set; since the question is chosen independently of document selection and answerability is nearly insensitive to question wording (\S~\ref{par:doc-verification}), we do not expect this to differentiate conditions.

\paragraph{Screened query set}\label{par:answer-screened} Document verification tests whether a strong reader generates the answer from the document, not whether the answer explicitly appears (string match). Some verified documents therefore contain the exact answer string and others do not. Copies inherit the anchor's status, but each added diverse document brings its own representation of relevant information. Thus improvements from the diverse condition could simply come from adding a document that contains a more explicit (or exact) wording of the answer, rather than any overall diverse benefit. To control for this we screen every document for the gold answer and remove two kinds of match: 
\begin{enumerate}
	\item A \emph{string match} occurs when the answer is present in the document as a contiguous string (case, punctuation, and whitespace normalized).
	\item  A \emph{near match} occurs when, after stop words are removed from answer and document, the answer's $n$ remaining words all appear, in any order and lightly stemmed, within a window of $2n{-}1$ document words. The window permits intervening words and catches restatements that reorder words, such as ``stabilize soil'' vs.\ ``soil stabilization''. 
\end{enumerate}
 This approach is conservative: coincidental proximity can cost us clean documents, but it never lets a match through. The results is our \emph{screened queryset}, containing 252 five-genre queries, rebuilt with the main construction, whose evidence carries no string or near match (Table~\ref{tab:dataset-properties}).

\begin{table}[t]
	\centering
	\caption{Properties of the extracted datasets. Documents are the unique fiction documents referenced by each queryset's verified orderings; document length is in whitespace words, gold-answer length in normalized tokens. The screened dataset additionally excludes near matches: a query counts at genre threshold $g$ only if its first $g$ ordering documents carry no near-match flag, and the document rows cover only unflagged documents.}
	\label{tab:dataset-properties}
	\begin{tabular}{l r r }
		\toprule
		& Main & Screened \\
		\midrule
		Queries (usable) & 2,756 & 1,443 \\
		\quad $\geq$2 genres & 2,736 & 888 \\
		\quad $\geq$3 genres & 2,718 & 644 \\
		\quad $\geq$4 genres & 2,381 & 440 \\
		\quad all 5 genres & 1,900 & 252 \\
		Documents & 1,492 & 1,237 \\
		\quad blog & 200 (13.4\%) & 187 (15.1\%) \\
		\quad corporate & 297 (19.9\%) & 239 (19.3\%) \\
		\quad encyclopedia & 200 (13.4\%) & 156 (12.6\%) \\
		\quad news & 495 (33.2\%) & 379 (30.6\%) \\
		\quad social & 300 (20.1\%) & 276 (22.3\%) \\
		Document length (words), max & 686 & 682 \\
		Document length (words), median & 507 & 506 \\
		Gold answer length (tokens), max & 10 & 10 \\
		Gold answer length (tokens), median & 2 & 3 \\
		\bottomrule
	\end{tabular}
\end{table}

\newtcolorbox{promptbox}[1][]{
	enhanced, sharp corners,
	colback=gray!5, colframe=gray!50, boxrule=0.5pt,
	fonttitle=\bfseries\small, coltitle=black,
	fontupper=\small\ttfamily,
	fontlower=\small\ttfamily,
	left=3pt, right=3pt, top=3pt, bottom=3pt, #1
}

\begin{figure}[t]
	\begin{promptbox}[title=Paragraph $\Rightarrow$ Paraphrase]
		Paraphrase the provided text, rephrasing the content in your own words while preserving all information. Include only information stated in the original text; do not add new facts or interpretations. Write only in English.
		
		\tcblower
		Text: \{document\}
		
		Paraphrase:
	\end{promptbox}
	\caption{Paraphrase transformation prompt. \texttt{\{document\}} is filled per document.}
	\label{fig:prompt-paraphrase}
\end{figure}

\subsection{Generation}

\paragraph{Generator Backbones} We evaluate four instruction-tuned generators spanning 1B--12B parameters across two model families: Llama-3.2-1B-Instruct, Llama-3.2-3B-Instruct, Llama-3.1-8B-Instruct, and Gemma-3-12B-IT. The range lets us observe whether the effects of redundant evidence vary with model size. 

GPT-4o-2024-08-06 verifies documents (see \emph{Document verification}), and Qwen2.5-32B-Instruct generates paraphrases and judges answer correctness (\S\ref{sec:evaluation}). 

\paragraph{Answer generation} For each query and condition, the generator receives the canonical question and the evidence set (duplicate, paraphrase or diverse), rendered as numbered \texttt{<document>} blocks in a fixed order, with an instruction to answer from the provided documents and to give the shortest correct answer (Figure~\ref{fig:prompt-rag}). The closed-book condition omits the documents and the reference to them (Figure~\ref{fig:prompt-closedbook}). Answers are decoded greedily (temperature 0) up to 64 new tokens.

\begin{figure}[t]
	\begin{promptbox}[title=Question $\Rightarrow$ Answer (closed-book)]
		Answer the question. Give the shortest correct answer possible. Output only the answer.
		
		\tcblower
		Question: \{question\}
		
		Answer:
	\end{promptbox}
	\caption{Closed-book generator prompt. \texttt{\{question\}} is filled per query.}
	\label{fig:prompt-closedbook}
\end{figure}

\begin{figure}[t]
	\begin{promptbox}[title=Documents + Question $\Rightarrow$ Answer (RAG)]
		Answer the question based on the provided documents. Give the shortest correct answer possible. Output only the answer.
		
		\tcblower
		Documents:\\
		\{context\}
		
		Question: \{query\}
		
		Answer:
	\end{promptbox}
	\caption{RAG generator prompt. \texttt{\{context\}} is the concatenated document representations in retrieval-rank order, separated by ``\texttt{\textbackslash n\textbackslash n-{}-{}-\textbackslash n\textbackslash n}''; \texttt{\{query\}} is the question.}
	\label{fig:prompt-rag}
\end{figure}

\subsection{Evaluation}
\label{sec:evaluation}

\paragraph{Final Answer Correctness} A generated answer is scored by an LLM judge (Qwen2.5-32B-Instruct, the same model that generates paraphrases), which compares it to the gold answer and returns CORRECT or INCORRECT at temperature 0 (Figure~\ref{fig:prompt-judge-accuracy}). Because the facts are fictional, the judge is instructed to compare only against the gold answer, not to rely on outside knowledge, and not to penalize unfamiliar entities. Equivalent forms, more specific answers, and answers that contain the gold answer count as correct; answers that omit it, contradict it, or refuse count as incorrect.

\begin{figure}[t]
	\begin{promptbox}[title=Generated + Gold $\Rightarrow$ Correct / Incorrect]
		You are an expert evaluator for a question-answering system. Compare the generated answer to the gold (correct) answer and decide whether the generated answer is correct.
		
		The questions concern FICTIONAL events, so judge ONLY by comparison to the gold answer. Do not rely on outside knowledge, and do not penalize an answer for naming unfamiliar entities.
		
		A generated answer is CORRECT if ANY of the following apply:
		- It conveys the same core information as the gold answer
		- It is a valid alternative form or spelling of the gold answer
		- It is more specific than the gold answer
		- It includes the gold answer along with additional information
		- It states the gold answer among a short list of items
		
		A generated answer is INCORRECT if ANY of the following apply:
		- It does not contain the gold answer or its equivalent anywhere in the response
		- It contradicts the gold answer
		- It is a non-answer, a refusal, or empty
		
		When in doubt, if the gold answer appears within the generated answer, respond CORRECT.
		
		Respond with ONLY one word: CORRECT or INCORRECT.
		
		\tcblower
		Question: \{query\}
		
		Generated Answer: \{generated\_answer\}
		
		Gold Answer(s):\\
		\{gold\_answer\}
		
		Evaluation (CORRECT or INCORRECT):
	\end{promptbox}
	\caption{LLM-judge accuracy evaluation prompt. \texttt{\{gold\_answer\}} is the set of accepted answers, each on its own line as a bulleted list; \texttt{\{query\}} and \texttt{\{generated\_answer\}} are filled per evaluation.}
	\label{fig:prompt-judge-accuracy}
\end{figure}

\paragraph{Metric validation.} We also compute exact match, token-level F1, and token recall (the fraction of gold-answer tokens appearing in the generated answer). Exact match is a strict lower bound on judged correctness: across all 171{,}228 scored answers, no exact-match answer is judged incorrect, and every disagreement (32.0\% of answers) is one the judge credits without a verbatim match. Token recall closely tracks the LLM judge across conditions, context sizes, and models; critically, both rise with $k$ in the diverse condition. Exact match and F1 instead drift downward as $k$ grows in every condition, while recall holds steady. The cause is answer length: answers lengthen as $k$ grows, and the models whose answers lengthen most show the largest exact-match decline. The drift therefore reflects longer answers, not less correct ones. 

Consequently we report LLM judged correctness as the headline metric. Full tables for all measures, including generated-answer lengths, are in the released repository.

\paragraph{Statistical analysis.} 
Statistical significance testing is done using McNemar tests ($p < 0.01$), comparing each condition against the single-source, $k=1$ baseline; all conditions are scored on the same 1,900 queries (\emph{Experimental sample} above), so any two conditions can be compared on identical queries. The repository's tables for F1, recall, and answer length mark significance the same way, using paired two-sided Wilcoxon signed-rank tests.

\section{Results}

%

\begin{figure*}
\vspace{10pt}
  \includegraphics[width=1.0\textwidth]{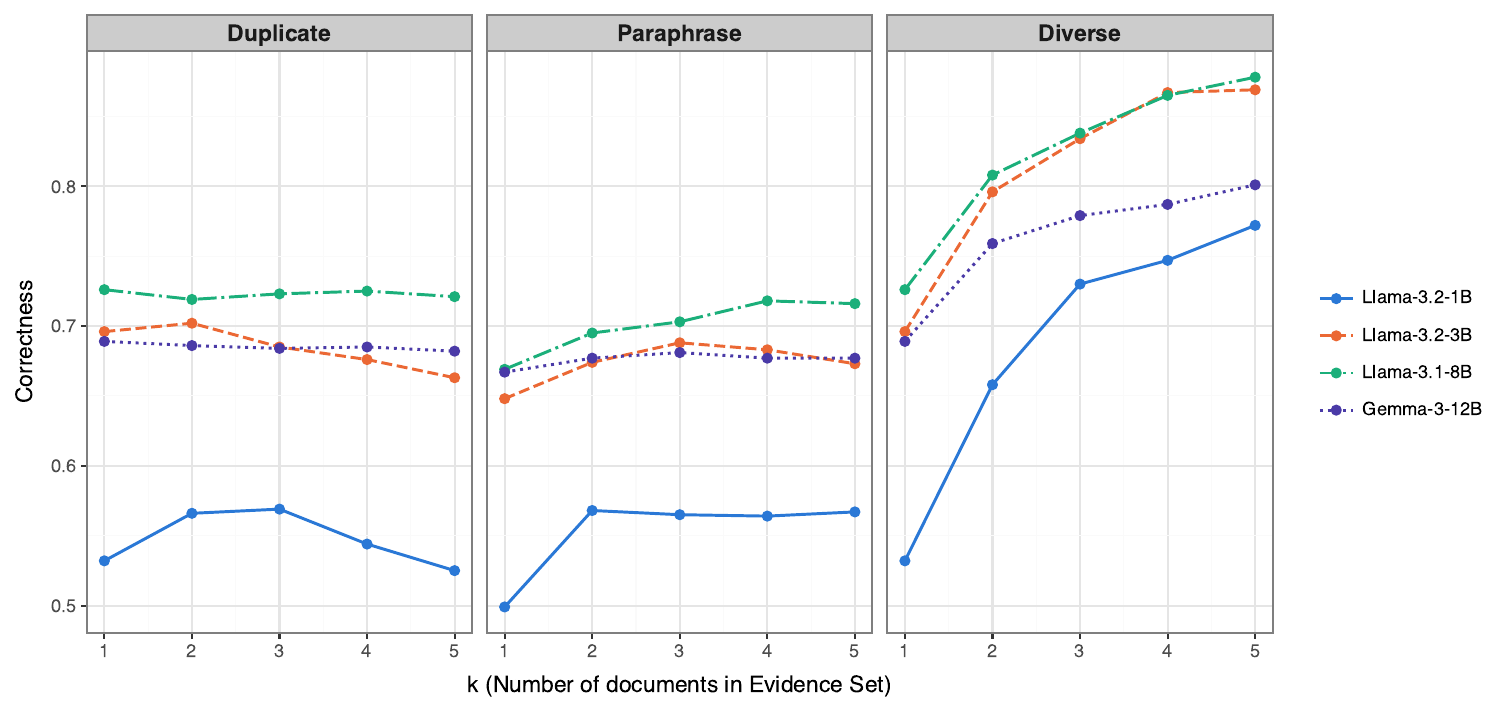}
  \caption{Visual summary of Table~\ref{tab:main_correctness} results. Note: $k=1$ represents the baseline single source (anchor doc) setting or the single source (paraphrased) setting.}
  \label{fig:correctness_by_k.pdf}
\end{figure*}

The main experimental results are provided in Table~\ref{tab:main_correctness}, which we reference throughout the results section. A visual summary is also provided in Figure~\ref{fig:correctness_by_k.pdf}. The results section is divided into 4 parts, each detailing a key finding from our experiments.

\begin{table*}[t]
	\centering
	\caption{LLM-judge correctness by condition, on the 1900 queries eligible at diverse depth $k{=}5$ (all conditions scored on the same queries). \emph{Single source} is the base document alone (duplicate and diverse coincide at $k{=}1$). $\blacktriangle$/$\blacktriangledown$: significantly above/below the Baseline (paired McNemar exact, $p{<}0.01$); no mark = not significant.}
	\label{tab:main_correctness}
	\begin{tabular}{l c rrrr}
		\toprule
		& & \multicolumn{4}{c}{G\bf enerator Backbone} \\[3pt]
		\cline{3-6} \\[-7pt]
		& \bf $k$ & Llama-3.2-1B & Llama-3.2-3B & Llama-3.1-8B & Gemma-3-12B \\
		\midrule
		\multicolumn{6}{l}{\emph{Baseline}} \\
		\rowcolor{gray!30}\quad Single source (anchor doc) & 1 & 0.532\phantom{\,$\blacktriangle$} & 0.696\phantom{\,$\blacktriangle$} & 0.726\phantom{\,$\blacktriangle$} & 0.689\phantom{\,$\blacktriangle$} \\
		\midrule
		\multicolumn{6}{l}{\emph{Reference}} \\
		\quad No evidence (closed-book) & 0 & 0.017\,$\blacktriangledown$ & 0.034\,$\blacktriangledown$ & 0.021\,$\blacktriangledown$ & 0.040\,$\blacktriangledown$ \\
		\quad Single source (paraphrased) & 1 & 0.499\,$\blacktriangledown$ & 0.648\,$\blacktriangledown$ & 0.669\,$\blacktriangledown$ & 0.667\,$\blacktriangledown$ \\
		\midrule
		\multicolumn{6}{l}{\emph{Evidence Set}} \\
		\quad Duplicate & 2 & 0.566\,$\blacktriangle$ & 0.702\phantom{\,$\blacktriangle$} & 0.719\phantom{\,$\blacktriangle$} & 0.686\phantom{\,$\blacktriangle$} \\
		\quad Paraphrase & 2 & 0.568\,$\blacktriangle$ & 0.674\phantom{\,$\blacktriangle$} & 0.695\,$\blacktriangledown$ & 0.677\phantom{\,$\blacktriangle$} \\
		\quad Diverse & 2 & 0.658\,$\blacktriangle$ & 0.796\,$\blacktriangle$ & 0.808\,$\blacktriangle$ & 0.759\,$\blacktriangle$ \\[4pt]
		
		\quad Duplicate& 3 & 0.569\,$\blacktriangle$ & 0.685\phantom{\,$\blacktriangle$} & 0.723\phantom{\,$\blacktriangle$} & 0.684\phantom{\,$\blacktriangle$} \\
		\quad Paraphrase& 3 & 0.565\,$\blacktriangle$ & 0.688\phantom{\,$\blacktriangle$} & 0.703\,$\blacktriangledown$ & 0.681\phantom{\,$\blacktriangle$} \\
		\quad Diverse& 3 & 0.730\,$\blacktriangle$ & 0.834\,$\blacktriangle$ & 0.838\,$\blacktriangle$ & 0.779\,$\blacktriangle$ \\[4pt]
		
		\quad Duplicate& 4 & 0.544\phantom{\,$\blacktriangle$} & 0.676\phantom{\,$\blacktriangle$} & 0.725\phantom{\,$\blacktriangle$} & 0.685\phantom{\,$\blacktriangle$} \\
		\quad Paraphrase& 4 & 0.564\,$\blacktriangle$ & 0.683\phantom{\,$\blacktriangle$} & 0.718\phantom{\,$\blacktriangle$} & 0.677\phantom{\,$\blacktriangle$} \\
		\quad Diverse& 4 & 0.747\,$\blacktriangle$ & 0.867\,$\blacktriangle$ & 0.865\,$\blacktriangle$ & 0.787\,$\blacktriangle$ \\[4pt]
		
		\quad Duplicate& 5 & 0.525\phantom{\,$\blacktriangle$} & 0.663\,$\blacktriangledown$ & 0.721\phantom{\,$\blacktriangle$} & 0.682\phantom{\,$\blacktriangle$} \\
		\quad Paraphrase& 5 & 0.567\,$\blacktriangle$ & 0.673\phantom{\,$\blacktriangle$} & 0.716\phantom{\,$\blacktriangle$} & 0.677\phantom{\,$\blacktriangle$} \\
		\quad Diverse& 5 & 0.772\,$\blacktriangle$ & 0.869\,$\blacktriangle$ & 0.878\,$\blacktriangle$ & 0.801\,$\blacktriangle$ \\
		\bottomrule
	\end{tabular}
\end{table*}

\subsection{Finding 1 - Redundancy and Paraphrasing have little impact of correctness}

Considering the Duplicate setting of Figure~\ref{fig:correctness_by_k.pdf}, noting that $k=1$ represents the baseline setting of a single document and $k=5$ represents 5 copies of the same document provided to the generator. 

Repeating content did not improve correctness. With duplicate copies, no generator differs significantly from its single-source baseline at any $k$, with two exceptions: the 1B generator gains 0.034--0.037 at $k{=}2$--$3$ but not beyond, and the 3B generator falls 0.033 below baseline at $k{=}5$. 

With paraphrased copies, only two differences are significant: the 1B generator gains 0.032--0.036 at every $k$, and the 8B generator falls below baseline at $k{=}2$--$3$. Overall though, repeating content via paraphrasing did not increase correctness.

\subsection{Finding 2 - Diverse documents improve correctness}

Adding diverse documents raises correctness significantly above baseline at every $k$ for every generator, and the gain rises monotonically with the number added: by $k{=}5$ it reaches 0.240 (1B), 0.173 (3B), 0.152 (8B), and 0.112 (12B).

At the same $k$, diverse documents outperform Duplicate and Paraphrase form. Diverse documents exceed Duplicate by 0.074--0.247 and Paraphrased copies by 0.083--0.205, significantly for every generator at every $k$, and the margin widens as $k$ grows. Paraphrased and Duplicate differ by at most 0.042, significantly in 3 of the 16 comparisons and in both directions.

\begin{table}[h!]
	\centering
	\caption{Matched-$k$ contrasts between the added-evidence conditions on LLM-judge correctness, on the 1900 queries eligible at diverse depth $k{=}5$. Each cell is the labeled improvement in correctness at equal evidence count $k$ (paired difference on the same queries; negative = a decline). $\blacktriangle$/$\blacktriangledown$: significantly above/below zero (paired McNemar exact, $p{<}0.01$); no mark = not significant.}
	\label{tab:condition-contrasts-main}

	\begin{tabularx}{1\columnwidth}{>{\raggedright\arraybackslash}p{7em} c XXXX}
		\toprule
		\emph{Improvement in judge correctness} & $k$ & Llama-3.2-1B & Llama-3.2-3B & Llama-3.1-8B & Gemma-3-12B \\
		\midrule
		Paraphrase over  & 2 & $+0.002$\phantom{\,$\blacktriangle$} & $-0.027$\,$\blacktriangledown$ & $-0.025$\,$\blacktriangledown$ & $-0.009$\phantom{\,$\blacktriangle$} \\
		Duplicate & 3 & $-0.004$\phantom{\,$\blacktriangle$} & $+0.003$\phantom{\,$\blacktriangle$} & $-0.021$\,$\blacktriangledown$ & $-0.003$\phantom{\,$\blacktriangle$} \\
		& 4 & $+0.021$\phantom{\,$\blacktriangle$} & $+0.007$\phantom{\,$\blacktriangle$} & $-0.007$\phantom{\,$\blacktriangle$} & $-0.007$\phantom{\,$\blacktriangle$} \\
		& 5 & $+0.042$\,$\blacktriangle$ & $+0.011$\phantom{\,$\blacktriangle$} & $-0.005$\phantom{\,$\blacktriangle$} & $-0.004$\phantom{\,$\blacktriangle$} \\
		\midrule
		Diverse over  & 2 & $+0.092$\,$\blacktriangle$ & $+0.095$\,$\blacktriangle$ & $+0.089$\,$\blacktriangle$ & $+0.074$\,$\blacktriangle$ \\
		Duplicate & 3 & $+0.161$\,$\blacktriangle$ & $+0.149$\,$\blacktriangle$ & $+0.115$\,$\blacktriangle$ & $+0.096$\,$\blacktriangle$ \\
		& 4 & $+0.204$\,$\blacktriangle$ & $+0.191$\,$\blacktriangle$ & $+0.141$\,$\blacktriangle$ & $+0.103$\,$\blacktriangle$ \\
		& 5 & $+0.247$\,$\blacktriangle$ & $+0.206$\,$\blacktriangle$ & $+0.157$\,$\blacktriangle$ & $+0.119$\,$\blacktriangle$ \\
		\midrule
		Diverse over  & 2 & $+0.091$\,$\blacktriangle$ & $+0.122$\,$\blacktriangle$ & $+0.114$\,$\blacktriangle$ & $+0.083$\,$\blacktriangle$ \\
		Paraphrase & 3 & $+0.165$\,$\blacktriangle$ & $+0.146$\,$\blacktriangle$ & $+0.136$\,$\blacktriangle$ & $+0.099$\,$\blacktriangle$ \\
		& 4 & $+0.183$\,$\blacktriangle$ & $+0.184$\,$\blacktriangle$ & $+0.147$\,$\blacktriangle$ & $+0.110$\,$\blacktriangle$ \\
		& 5 & $+0.205$\,$\blacktriangle$ & $+0.196$\,$\blacktriangle$ & $+0.162$\,$\blacktriangle$ & $+0.123$\,$\blacktriangle$ \\
		\bottomrule
	\end{tabularx}
\end{table}

\subsection{Finding 3 - Diverse gain survives when the answer string is absent}
\label{sec:answer-screened}

The Finding 2 results show that Diverse improves correctness; this could in principle come from adding additional documents that contain the explicit answer string. The screened queryset removes this situation by removing any document that explicitly contain the answer (\S\ref{par:answer-screened}). We then rescore the 8B generator on its 252 five-genre queries. Table~\ref{tab:condition-correctness-ablation} presents these results.

The Diverse gain survives. Correctness under Diverse documents rises monotonically from 0.496 at $k{=}2$ to 0.647 at $k{=}5$, significantly above the 0.417 baseline from $k{=}3$ onward (gains of 0.139--0.230). Instead, the Duplicate and Paraphrase do not differ significantly from baseline at any $k$. At the same $k$, the pattern repeats (Table~\ref{tab:condition-contrasts-ablation}). Diverse documents exceed Duplicate significantly at every $k$, by 0.087--0.218, and Paraphrased  from $k{=}3$, by 0.111--0.164. Paraphrased and Duplicate never differ significantly.

We conclude that the benefit from giving the generator diverse documents comes truely from their diversity in genre and not from having more or better relevant information.

\begin{table}[h!]
	\centering
	\caption{LLM-judge correctness by condition, on the 252 queries eligible at diverse depth $k{=}5$ (all conditions scored on the same queries). Screened queryset (\S\ref{par:answer-screened}). \emph{Single source} is the base document alone (duplicate and diverse coincide at $k{=}1$). $\blacktriangle$/$\blacktriangledown$: significantly above/below the Baseline (paired McNemar exact, $p{<}0.01$); no mark = not significant.}
	\label{tab:condition-correctness-ablation}
	\begin{tabular}{l c r}
		\toprule
		& $k$ & Llama-3.1-8B \\
		\midrule
		\multicolumn{3}{l}{\emph{Baseline}} \\
		\rowcolor{gray!30}\quad Single source (base doc) & 1 & 0.417\phantom{\,$\blacktriangle$} \\
		\midrule
		\multicolumn{3}{l}{\emph{Reference}} \\
		\quad No evidence (closed-book) & 0 & 0.008\,$\blacktriangledown$ \\
		\quad Single source (paraphrased) & 1 & 0.347\phantom{\,$\blacktriangle$} \\
		\midrule
		\multicolumn{3}{l}{\emph{Added evidence}} \\
		\quad Duplicate & 2 & 0.409\phantom{\,$\blacktriangle$} \\
		\quad Paraphrase & 2 & 0.409\phantom{\,$\blacktriangle$} \\
		\quad Diverse & 2 & 0.496\phantom{\,$\blacktriangle$} \\[4pt]

		\quad Duplicate & 3 & 0.417\phantom{\,$\blacktriangle$} \\
		\quad Paraphrase & 3 & 0.440\phantom{\,$\blacktriangle$} \\
		\quad Diverse & 3 & 0.556\,$\blacktriangle$ \\[4pt]
				
		\quad Duplicate & 4 & 0.421\phantom{\,$\blacktriangle$} \\
		\quad Paraphrase & 4 & 0.444\phantom{\,$\blacktriangle$} \\
		\quad Diverse & 4 & 0.607\,$\blacktriangle$ \\[4pt]
		
		\quad Duplicate & 5 & 0.429\phantom{\,$\blacktriangle$} \\		
		\quad Paraphrase & 5 & 0.476\phantom{\,$\blacktriangle$} \\		
		\quad Diverse & 5 & 0.647\,$\blacktriangle$ \\
		\bottomrule
	\end{tabular}
\end{table}

\begin{table}[h!]
	\centering
	\caption{Matched-$k$ contrasts between the added-evidence conditions on LLM-judge correctness, on the 252 queries eligible at diverse depth $k{=}5$. Screened queryset (\S\ref{par:answer-screened}). Each cell is the labeled improvement in correctness at equal evidence count $k$ (paired difference on the queries carrying both conditions; a row label shows the pair count when it differs from the balanced set; negative = a decline). $\blacktriangle$/$\blacktriangledown$: significantly above/below zero (paired McNemar exact, $p{<}0.01$); no mark = not significant.}
	\label{tab:condition-contrasts-ablation}
	\begin{tabular}{l c r}
		\toprule
		\emph{Improvement in judge correctness} & $k$ & Llama-3.1-8B \\
		\midrule
		\quad Paraphrase over Duplicate (n=225) & 2 & $-0.009$\phantom{\,$\blacktriangle$} \\
		& 3 & $+0.018$\phantom{\,$\blacktriangle$} \\
		& 4 & $+0.009$\phantom{\,$\blacktriangle$} \\
		& 5 & $+0.031$\phantom{\,$\blacktriangle$} \\
		\midrule
		\quad Diverse over Duplicate & 2 & $+0.087$\,$\blacktriangle$ \\
		& 3 & $+0.139$\,$\blacktriangle$ \\
		& 4 & $+0.187$\,$\blacktriangle$ \\
		& 5 & $+0.218$\,$\blacktriangle$ \\
		\midrule
		\quad Diverse over Paraphrase (n=225) & 2 & $+0.084$\phantom{\,$\blacktriangle$} \\
		& 3 & $+0.111$\,$\blacktriangle$ \\
		& 4 & $+0.160$\,$\blacktriangle$ \\
		& 5 & $+0.164$\,$\blacktriangle$ \\
		\bottomrule
	\end{tabular}
\end{table}

\section{Finding 4 - Document genres influence correctness}

Recall that documents in the FictionalQA dataset fit within 5 genres: news, social media, corporate, encyclopedia and blogs. As an ablation we consider the impact of when the anchor document $k_1$ comes from different genres. Table ~\ref{tab:base-genre-correctness} shows this result. 

The choice of genre heavily influences correctness. Encyclopaedic documents drive the highest correctness (0.701--0.846). Social media documents drive the lowest correctness (0.372--0.619). Recall that all documents were verified to contain information relevant to generate a correct answer; so the effect of correctness is not that the document does not contain relevant information but instead the genre itself that influences correctness. 

\begin{table}[t]
  \centering
  \caption{LLM-judge correctness with the base document alone (duplicate\_k1), by base-document genre (stratified $\sim$20\% each, full main queryset). The last row gives the per-model $\chi^2$ test of independence between base genre and correctness (dof=4). Pooled = all models' verdicts combined.}
  \label{tab:base-genre-correctness}
  \begin{tabularx}{1\columnwidth}{l XXX XXX }
    \toprule
    Genre & $n$ & Llama-3.2-1B & Llama-3.2-3B & Llama-3.1-8B & Gemma-3-12B & Pooled \\
    \midrule
    blog & 552 & 0.562 & 0.764 & 0.770 & 0.748 & 0.711 \\
    corp. & 551 & 0.526 & 0.650 & 0.697 & 0.664 & 0.634 \\
    ency. & 551 & 0.701 & 0.844 & 0.846 & 0.824 & 0.804 \\
    news & 551 & 0.561 & 0.735 & 0.766 & 0.726 & 0.697 \\
    social & 551 & 0.372 & 0.555 & 0.619 & 0.581 & 0.532 \\
    \midrule
    all & 2756 & 0.544 & 0.710 & 0.739 & 0.709 & 0.676 \\
    $\chi^2$ $p$ & & 1.9e-25 & 2.1e-27 & 2.8e-17 & 1.8e-18 & 8.5e-87 \\
    \bottomrule
  \end{tabularx}
\end{table}

\section{Discussion \& Conclusion}

Through a series of controlled experiments we have investigate the role of redundant and diverse information a retriever might provide to a generator in a RAG setting. While classic Information Retrieval systems often aim to reduce redundancy and increase diversity to suit human preferences~\cite{santos2015search}, it's not clear whether this should be the same for generators.

To answer this question we control retrieval with $k=1-5$ documents according to three scenarios: 1) $k$ duplicate documents provided to the generator; 2) $k$ LLM-paraphrased documents provided to the generator; and 3) $k$ diverse document same of five genres (news, blogs, etc) provided to the generator. All documents contain information relevant to the question. The FictionalQA dataset means controls for parametric generator knowledge (as confirmed by near zero close-book correctness).

Evaluating open-source models across four parameter scales ($1\text{B}$--$12\text{B}$), we isolated the mechanistic impact of context redundancy versus diversity. Our experiments demonstrate that:
\begin{itemize}
    \item \textbf{Pure redundancy or paraphrasing offers minimal utility:} Supplying duplicate copies or LLM-generated paraphrases of an answer-bearing document fails to meaningfully improve answer correctness over a single-document baseline. This is somewhat contrary to previous studies that show redundancy paraphrasing may help.
    \item \textbf{Diverse evidence sets provide substantial gains:} Incorporating diverse documents spanning multiple genres increases correctness by up to $24.7$ percentage points over duplicate contexts. Controlling for explicit string matches confirms that this gain stems from diverse genres rather than the presence of exact match answer phrasing.
    \item \textbf{Genre composition matters:} The intrinsic genre of retrieved documents significantly influences model output, with encyclopedic contexts consistently outperforming social media genres even when answerability is held constant.
\end{itemize}

Our results point to a rethink of how RAG retrieval is done. Rather than filling $k$-document context windows with top-$k$ passages that contain near-identical phrasing, retrievers and rerankers should explicitly select for diverse documents and independent perspectives. Bridging the gap between retriever scoring and generator context preference offers a promising direction for future research in agentic search.

Code and results are provided in \github.
%

\bibliographystyle{ACM-Reference-Format}
\bibliography{bibliography/references}

\end{document}